\documentclass[12pt]{article}
\usepackage{amsmath,amsfonts,amssymb}
\usepackage{hyperref}
\usepackage{graphicx}
\usepackage{slashed}
\usepackage{color} 
\usepackage{epsfig}
\usepackage{psfrag}
\begin{document}
\bibliographystyle{utphys}

\hfill {\tt UT-Komaba/15-5}

\vspace{6mm}

\begin{center}
{\LARGE Mirror symmetry and the flavor vortex operator }\\
\vskip 3mm
{\LARGE  in two dimensions}
\end{center}
\vspace{3mm}

\begin{center}
Takuya Okuda\\
\vspace{3mm}
University of Tokyo \\
Komaba, Meguro-ku, Tokyo 153-8902, Japan
\end{center}

\vskip 5mm

\abstract{
\vskip 2mm
The flavor vortex operator $V_\alpha$ is a local disorder operator defined by coupling a two-dimensional $\mathcal{N}=(2,2)$ chiral multiplet to a non-dynamical gauge field with vortex singularity of holonomy $2\pi\alpha$.
We show that it is related to the mirror-dual twisted chiral multiplet, with bottom component $y$, as~$V_\alpha=e^{-\alpha y}$.
}

\section{Introduction}
\label{sec:introduction}

In quantum field theories there exist operators that are defined by singular boundary conditions imposed on dynamical fields in the path integral.
Equivalently, such operators can be defined through the coupling to a singular background field, by which the dynamical field is shifted.
It is also meaningful to define an operator by coupling the theory to a singular background field even when there is no dynamical field to be shifted.
Much progress has been made in recent years in understanding operators of these kinds, the so-called disorder operators.
Duality often exchanges disorder operators with more conventional operators defined as functions of dynamical fields.
In supersymmetric field theories, some correlation functions of disorder operators preserving part of supersymmetry can be exactly computed by localization.

In this note we study the $\mathcal{N}=(2,2)$ theory on Euclidean plane consisting of a single chiral multiplet without a superpotential.
A ``local'' disorder operator, the flavor vortex operator, can be defined by coupling the theory to a background gauge field for the flavor $U(1)$ symmetry of the form
\begin{equation} \label{vortex-singularity}
v_\mu dx^\mu \sim \alpha d\varphi\,,
\end{equation}
where $\varphi$ is the angular coordinate around the insertion point and $\alpha$ a real parameter.
By also turning on the background auxiliary field $D$ in the non-dynamical vector multiplet, half of the supercharges can be preserved.
The analysis of broken supercharges will reveal the connection to mirror symmetry \cite{Hori:2000kt}.

A closely related paper \cite{Hosomichi:2015pia} studying gauge vortex operators appeared recently.

\section{Flavor vortex operator}
\label{sec:flav-vort-oper}

A chiral multiplet of two-dimensional $\mathcal{N}=(2,2)$ supersymmetry consists of a complex scalar $\phi$, fermions $\psi_\mp$, and a complex auxiliary field $F$.  
We put the theory on Euclidean plane with metric $ds^2=|dz|^2$, where $z=x^1+ix^2$.
Their transformations under the supercharges $(Q_\pm, \overline{Q}_\pm)$ are given in (\ref{SUSY-chiral}).
The free theory of a single chiral multiplet without a superpotential, possibly with a twisted mass, has a $U(1)$ flavor symmetry.
We let the fields in the chiral multiplet have charge $+1$.
Fields $(\sigma, \overline{\sigma}, v_\mu, \lambda_\pm, \overline{\lambda}_\pm, D)$ in a vector multiplet transform according to (\ref{SUSY-vector}).
We are interested in configurations of a non-dynamical vector multiplet for which the transformations under some supercharges vanish.
The twisted mass is simply the constant vev of the complex scalar $\sigma$.

It is cleaner to start with a smooth disorder operator of finite size, as in references \cite{Kapustin:2012iw,Hook:2013yda} that studied the three-dimensional case.
It is defined by coupling the chiral multiplet to a smooth background gauge field $v=v_\mu dx^\mu$.
The configuration preserves those supercharges for which the variation of the fermions in the flavor gauge multiplet vanishes.
From (\ref{SUSY-vector})  we see that  $Q_-$ and $\overline{Q}_+$ are preserved when the condition~$D =2i v_{z\overline{z}}$, or equivalently $D dx^1\wedge dx^2+dv=0$, is satisfied.
(Similarly $Q_+$ and $\overline{Q}_-$ are preserved when  $D = -2i v_{z\overline{z}}$.)
We characterize the field configuration by a real function $\rho=\rho(x)$, the ``vorticity density'', with support in a compact region:
\begin{equation}
D=-2\pi \rho\,,\qquad v_{z\overline{z}} = \pi i \rho \,.  
\end{equation}
Other fields in the vector multiplet are set to zero.
This defines a disorder operator; let us call it the smeared vortex operator~$V[\rho]$.

For the action $S=\int d^2x \mathcal L,$ the Lagrangian
\begin{equation}\label{Lag}
  \begin{aligned}
  \mathcal L&=  D^\mu\overline\phi D_\mu\phi +\overline\sigma  \sigma  \overline\phi\phi - \overline{\phi} D\phi  -\overline F F 
\\
&\qquad \qquad 
+2i \overline\psi{}_+ D_{ z} \psi_+ -2 i \overline\psi{}_- D_{ {\overline{z}}} \psi_-  +\overline\sigma \overline\psi{}_+\psi_-  +\sigma \overline\psi_- \psi_+ \,,
\end{aligned}
\end{equation}
and the transformations (\ref{SUSY-chiral}), the supersymmetry current $S^\mu$ corresponding to the linear combination $\delta= i\epsilon_+ Q_- - i\epsilon_- Q_+  -i \overline\epsilon{}_+ \overline Q{}_- + i\overline \epsilon{}_- \overline Q{}_+$ of supercharges is given as
\begin{equation} \label{SUSY-current-chiral}
  \begin{aligned}
  S^z =   -4 \epsilon _- (D_{\overline{z}} \bar{\phi })\psi _++4 \bar{\epsilon }_-   (D_{\overline{z}}\phi )\bar{\psi }_++2 i \sigma  \epsilon _+ \bar{\phi} \psi _++2 i \bar{\sigma } \bar{\epsilon }_+ \phi \bar{\psi }_+  \,,
\\
  S^{\overline{z}} = + 4 \epsilon _+ (D_z \bar{\phi })\psi _--4 \bar{\epsilon }_+ (D_z \phi) \bar{\psi }_-+2 i \epsilon _-   \bar{\sigma } \bar{\phi }\psi _-+2 i \sigma  \bar{\epsilon }_- \phi \bar{\psi }_-  \,.
  \end{aligned}
\end{equation}
Its derivation is reviewed in Appendix \ref{ST}.
Let $n_\mu$ be the outward normal vector to the boundary of the support of $\rho$, and $ds$ the line element.
By the standard argument \cite{Pol-Ch2} the integral 
\begin{equation} \label{contour-integral-normal-vector}
-  \oint ds\, n_\mu S^\mu = -  \int d^2x\, \partial_\mu S^\mu
\end{equation}
in the defining background for $V[\rho]$ equals the variation $\delta V[\rho]$.
A straightforward computation gives
\begin{equation}
\delta  V[\rho] =   4\pi  \int  d^2x \, \rho(x) ( \epsilon_- \overline{\phi}\,\psi_+ + \overline\epsilon_+ \overline{\psi}_- \phi)  \cdot V[\rho]\,.
\end{equation}
Terms proportional to $\epsilon_+$ and $\overline{\epsilon}_-$ have dropped out because $Q_-$ and $\overline{Q}_+$ are preserved.

We now take the limit
\begin{equation}
\rho \rightarrow  \alpha\cdot  \delta^2(x)\,,
\end{equation}
in which the smeared operator becomes a local operator inserted at the origin: $V[\rho]\rightarrow V_\alpha$.
It is the two-dimensional analog of the flavor vortex loop operator in three dimensions studied in \cite{MS,Kapustin:2012iw,Drukker:2012sr}.
(See also \cite{Yoshida:2014ssa,Assel:2015oxa,Beasley:2014jta} for the study of the same and the related operators in supersymmetric and non-supersymmetric settings.)

When $\alpha$ takes a non-integer real value, the parallel transport of matter fields around $V_\alpha$ produces a non-trivial phase; the fields and $V_\alpha$ are not mutually local.
In the terminology of \cite{Kapustin:2014gua,Aharony:2013hda,Gaiotto:2014kfa}, $V_\alpha$ is not a genuine local operator, but is an end of a topological line operator.

For the supersymmetry variations we get
\begin{equation}
 \overline{Q}_-\cdot V_\alpha = 4 \pi i \alpha\,    \overline{\psi }_- \phi\cdot  V_\alpha \,,\qquad
 Q_+\cdot  V_\alpha = 4 \pi i \alpha\, \overline{\phi}\, \psi_+ \cdot V_\alpha  \,.
\end{equation}
We see that the variations are proportional to $\alpha$ and $V_\alpha$.
This makes us propose that $V_\alpha$ can be written as
\begin{equation} \label{V-y}
  V_\alpha = e^{-\alpha y}\,,
\end{equation}
where the local operator $y$ is annihilated by $Q_-$ and $\overline{Q}_+$ and hence is a twisted chiral operator.
If we set
\begin{equation}\label{Y-Phi-relation-one}
\overline{\chi}_+ := i Q_+ y =  4 \pi \overline{\phi} \,\psi_+\,,
\quad  
\chi_- :=- i \overline Q_- y = - 4 \pi \overline{\psi}_- \phi\,,
\quad  
E := - 4 \pi \overline\psi_- \psi_+ \,,
\end{equation}
the fields $(y, \chi_-, \overline\chi_+, E)$ form a twisted chiral multiplet transforming as in (\ref{SUSY-twisted-chiral}).

The procedure reviewed in Appendix \ref{ST} gives the energy-momentum tensor
\begin{equation}\label{energy-momentum}
  \begin{aligned}
  T_{zz} &=   -4\pi D_z \overline{\phi} D_z \phi+\pi i   \overline{\psi}_- D_z \psi_- -\pi i   (D_z \overline{\psi}_-)\psi_-  \,,
\\
  T_{\overline{z}\overline{z}} &=   -4\pi D_{\overline{z}} \overline{\phi} D_{\overline{z}} \phi -\pi i \overline{\psi}_+ D_{\overline{z}} \psi_+ + \pi i (D_{\overline{z}} \overline{\psi}_+)\psi_+  \,,
\\
  T_{z\overline{z}} &= \pi |\sigma|^2 |\phi|^2 - \pi D |\phi|^2 + \frac{\pi }{2} \overline{\psi}_+\overline{\sigma}\psi_- + \frac{\pi }{2}\overline{\psi}_- \sigma \psi_+ \,.
  \end{aligned}
\end{equation}
This time we can use the Ward identity \cite{Pol-Ch2}
\begin{equation}
\frac{\partial}{\partial x^\mu} \mathcal O(x) = \frac{1}{2\pi} \int d^2x\, \partial_\nu T^\nu{}_\mu \cdot \mathcal O(x)
\end{equation}
to compute the derivatives $\partial_\mu V_\alpha$.  In terms of $y$, the result is
\begin{equation}\label{Y-Phi-relation-two}
  \partial_z y =4\pi ( D_z  \overline{\phi} )\phi - 2\pi i \overline{\psi}_- \psi_- \,, \qquad 
  \partial_{\overline{z}} y =  4\pi   \overline{\phi} D_{\overline{z}} \phi  - 2\pi i \overline{\psi}_+ \psi_+ \,.
\end{equation}

Using $y$ we can also write the smeared vortex operator as
\begin{equation}
V[\rho] = \exp\left( \int d^2x \rho(x) y(x) \right)\,.
\end{equation}
 
\section{Mirror symmetry}
\label{sec:conn-mirr-symm}

The Hori-Vafa mirror symmetry \cite{Hori:2000kt} specialized to a single chiral multiplet is an abelian duality applied to the phase of the chiral superfield $\Phi = \phi+\theta^-\psi_-+\theta^+\psi_+ +\theta^+ \theta^- F + \ldots$.
The Landau-Ginzburg model mirror to the original theory contains a twisted chiral superfield $Y$, and is characterized by the twisted superpotential
\begin{equation}
  \widetilde{W} = \sigma Y + \mu\, e^{-Y}\,,
\end{equation}
where $\mu$ is a mass scale.
The superfield $Y$ and its conjugate $\overline{Y}$ are related to $\Phi$, its conjugate $\overline{\Phi}$, and the background vector superfield $\mathcal{V}$ as
\begin{equation}
  Y + \overline{Y} = 4 \pi \overline{\Phi} e^{\mathcal{V}} \Phi \,,
\end{equation}
up to an additive renormalization of ${\rm Re}\,y$.
Expansion in $\theta^\mp$ and $\overline{\theta}^\mp$ gives relations among the fields in the dual theories~\cite{Hori:2000kt}.
These include precisely (\ref{Y-Phi-relation-one}) and (\ref{Y-Phi-relation-two}), if $Y$ is related to $(y,\chi_-,\overline{\chi}_+,E)$ in (\ref{V-y}) and (\ref{Y-Phi-relation-one}) as
\begin{equation} \label{Y-y}
Y=y+\overline{\theta}{}^-\chi_- + \theta^+ \overline{\chi}{}_++ \theta^+\overline{\theta}{}^- E + \ldots\,.
\end{equation}
We thus claim that under mirror symmetry, the flavor vortex operator $V_\alpha=e^{-\alpha y}$ and the twisted chiral superfield $Y$ in the mirror theory are related as~(\ref{Y-y}).
  
The imaginary part of $y$ in the mirror theory is a compact boson with identification ${\rm Im}\, y\sim {\rm Im}\, y+2\pi $.
Thus the exponential $ e^{  - \alpha y}$ for a generic real value of $\alpha$ is ill-defined as a local operator, as we found in the original theory.
Such an exponential operator with non-integer $\alpha$ was considered in \cite{Cecotti:2010fi,Cecotti:2013mba}.

More quantitative results and cases with gauge symmetries will be discussed elsewhere~\cite{with:Kazuo}.

\section*{Acknowledgements}

The author thanks Kazuo Hosomichi for useful discussions.
He is also grateful to the organizers and the participants of the workshop ``Gauge theories, supergravity and superstrings'' at Centro de Ciencias de Benasque Pedro Pascual for providing a stimulating atmosphere.
 This research is supported in part by Grant-in-Aid for Scientific Research~(B)~No.~25287049.

\appendix

\section{Supersymmetry transformations}
\label{sec:susy-transformations}

Our conventions are similar to those of \cite{MR2003030}.
We parametrize the supersymmetry transformation on Euclidean plane by Grassmann even parameters $(\epsilon_\pm, \overline{\epsilon}_\pm)$ as
\begin{equation}
\delta = i\epsilon_+ Q_- - i\epsilon_- Q_+  -i \overline\epsilon_+ \overline Q_- + i\overline \epsilon_- \overline Q_+\,.
\end{equation}
The gauge field $v_\mu$ appears in the covariant derivative as $D_\mu=\partial_\mu + i n v_\mu$ for a field of charge $n$.
Fields in chiral and anti-chiral multiplets transform as
\begin{equation}\label{SUSY-chiral}
  \begin{aligned}
&
\hspace{10mm}
\delta\phi=   + \epsilon_+\psi_--\epsilon_-\psi_+\,,
\qquad
 \delta\overline\phi= -\overline   \epsilon_+\overline \psi_- + \overline \epsilon_- \overline \psi_+\,,
 \\
&
\hspace{3mm}
\delta\psi_-=\epsilon_- F + \overline{\epsilon}_-\sigma \phi +2i\overline{\epsilon}_+ D_{z}\phi\,,
\qquad
\delta\psi_+=\epsilon_+ F - \overline{\epsilon}_+ \overline\sigma \phi +2i\overline{\epsilon}_- D_{\overline z}\phi\,,
\\
&
\hspace{3mm}
 \delta\overline\psi_-=\overline \epsilon_- \overline F + \epsilon_- \overline \sigma \overline\phi - 2i \epsilon_+ D_{ z}\overline \phi\,,
\qquad
 \delta\overline\psi_+=\overline \epsilon_+ \overline F - \epsilon_+ \sigma \overline\phi - 2i \epsilon_- D_{\overline z}\overline \phi\,,
 \\
&
\delta F=+\overline\sigma\overline\epsilon_+\psi_- + \sigma\overline{\epsilon}_- \psi_+ 
+ 2i  \overline{\epsilon}_+ D_{ z}\psi_+ - 2i \overline{\epsilon}_-  D_{\overline z}\psi_- -i (\overline{\epsilon}_+ \overline{\lambda}_--\overline{\epsilon}_- \overline{\lambda}_+)\phi\,.
\\
& \delta \overline{F} = -\sigma\epsilon_+ \overline\psi_- - \overline\sigma \epsilon_-\overline\psi_+
  +2i  \epsilon_+ D_{ z}\overline\psi_+ - 2i  \epsilon_- D_{\overline z}\overline{\psi}_- -i (\epsilon_+\lambda_- - \epsilon_- \lambda_+)\overline\phi\,.
  \end{aligned}
\end{equation}
Similarly, for an abelian vector multiplet in the Wess-Zumino gauge we have
\begin{equation} \label{SUSY-vector}
  \begin{aligned}
&
\hskip 10mm
\delta v_{ z}  = - \frac{ i }{ 2 } \epsilon_- \overline{\lambda}_-  - \frac{ i }{ 2 } \overline{\epsilon}_- \lambda_- \,, \qquad
\delta v_{\overline z}  = + \frac{ i }{ 2 } \epsilon_+ \overline{\lambda}_+  + \frac{ i }{ 2 } \overline{\epsilon}_+ \lambda_+ \,,
\\
&
\hskip 13mm
\delta \sigma  = - i \epsilon_- \overline{\lambda}_+ - i \overline{\epsilon}_{+} \lambda_- \,, \qquad 
\delta \overline{\sigma} = - i \overline{\epsilon}_- \lambda_+ - i \epsilon_+ \overline{\lambda}_- \,,
\\
&
\delta\lambda_-=+ i\epsilon_- \left(  D+2i v_{ z {\overline{z}}} \right) -2\epsilon_+ \partial_{ z}\sigma\,,
\qquad
    \delta\lambda_+= +i\epsilon_+ \left( D -2i v_{ z {\overline{z}}} \right) + 2\epsilon_- \partial_{{\overline z}}\overline\sigma\,,
\\
&    \delta\overline\lambda_-=- i\overline\epsilon_- \left( D - 2i v_{ z {\overline{z}}} \right)
-2\overline\epsilon_+ \partial_{ z}\overline\sigma\,, \qquad
    \delta\overline\lambda_+= - i\overline\epsilon_+ \left(  D + 2i v_{ z {\overline{z}}} \right) + 2\overline \epsilon_- \partial_{{\overline z}}\sigma\,,
\\
&
\hskip 27mm
\delta D  =\partial_{ z}( \overline{\epsilon}_+ \lambda_+ - \epsilon_+ \overline{\lambda}_+ ) - \partial_{ {\overline{z}}}( \overline{\epsilon}_-\lambda_- - \epsilon_- \overline{\lambda}_-) 
\,. 
  \end{aligned}
\end{equation}
Fields in twisted chiral and twisted anti-chiral multiplets transform as
\begin{equation}\label{SUSY-twisted-chiral}
  \begin{aligned}
&
\hskip 5mm
 \delta y = \overline{\epsilon}_+ \chi_- -\epsilon_- \overline{\chi}_+\,, \qquad   \delta \overline{y}  = - \epsilon_+ \overline{\chi}_- + \overline{\epsilon}_- \chi_+ \,, 
 \\
&
\hskip 2mm
\delta\chi_- = + 2i \epsilon_+ \partial_{{z}} y +\epsilon_- E\,, \qquad \delta \overline{\chi}_+ = 2i \overline{\epsilon}_- \partial_{{\overline{z}}}y + \overline{\epsilon}_+ E\,,
\\
&
\hskip 2mm
  \delta \overline{\chi}_- = - 2i \overline{\epsilon}_+ \partial_{{z}} \overline{y} + \overline{\epsilon}_- \overline{E} \,,  \qquad
 \delta \chi_+  = - 2i \epsilon_- \partial_{{\overline{z}}} \overline{y} + \epsilon_+ \overline{E} \,, 
\\
&
\hskip -2mm
\delta E = 2i \epsilon_+ \partial_{ z} \overline{\chi}_+ -2i \overline{\epsilon}_- \partial_{{\overline{z}}} \chi_-\,,\qquad 
  \delta \overline{E}  = 2i \overline{\epsilon}_+ \partial_{{z}} \chi_+ - 2i \epsilon_- \partial_{{\overline{z}}} \overline{\chi}_- \,.
  \end{aligned}
\end{equation}

\section{Derivation of $S^\mu$ and $T_{\mu\nu}$}
\label{ST}

The Noether current $S^\mu$ for the supersymmetry transformation $\delta \Phi^I$ is 
\begin{equation}
S^\mu=\delta\Phi^I [\partial (\partial_\mu \Phi^I) \backslash \partial \mathcal{ L } ] - K^\mu  \,,
\end{equation}
where $K^\mu$ is defined by $\delta \mathcal L = \partial_\mu K^\mu$ and $\partial x\backslash \partial f$ denotes the left derivative.
To compute $K^\mu$ we note that the Lagrangian $\mathcal L$ is, up to total derivatives, the $D$-component of the general supermultiplet with bottom component $\frac{1}{2} \overline{\phi}\phi$.
The transformation of the general multiplet allows us to express $\mathcal{L}$ as a total derivative and leads to~(\ref{SUSY-current-chiral}).

Let us use symbols with and without hats to denote frame and coordinate indices, respectively.
The Lagrangian (\ref{Lag}) can be covariantized in a natural way when the theory is coupled to the background metric $h_{\mu\nu}$.
To compute~$T_{\mu\nu}=-4\pi  h^{-1/2} \delta S/\delta h^{\mu\nu}$, we need the variation of the spin connection $ \omega_{\mu\hat{\nu}\hat{\rho}} $ that appears in the spinor covariant derivative $  D_ \mu =\partial_\mu +i v_\mu +\frac{1}{2}\omega_{\mu\hat{\nu}\hat{\rho}} [\gamma^{\hat{\nu}}, \gamma^{\hat{\rho}}  ]$.
Requiring that the vielbein~$e_{\hat{\mu}}{}^\nu$ is covariantly constant, we obtain (see for example \cite{Lorenzen:2014pna})
\begin{equation}
  \delta \omega_{\mu\hat{\nu}\hat{\rho}} =   \nabla_\nu\left( h_{\mu\lambda} e_{[\hat{\nu}}{}^\nu e_{\hat{\rho}]}{}_\sigma \delta h^{\lambda\sigma} + \frac{1}{2} e_{\hat{\nu} \lambda} e_{\hat{\rho} \sigma} \delta^\nu_\mu \delta h^{\lambda\sigma} - e_{\hat{\rho}\sigma} \delta^\nu_\mu \delta e_{\hat{\nu}}{}^\sigma
\right)\,.
\end{equation}
Applying this to the computation of $\delta S$, one finds that the terms containing $\delta e_{\hat{\mu}}{}^\nu$ cancel out.
A straightforward and as usual tedious computation yields the result (\ref{energy-momentum}).

\bibliography{refs}

\end{document}